\documentclass[12pt]{article}
\usepackage{epsf}
\usepackage{slashed}
\usepackage{color}
\usepackage{hyperref}
\begin{document}
\begin{titlepage}
\title{MOND and Natural Scales of Distance and Mass}
\author{
{Piotr \.Zenczykowski }\footnote{E-mail: piotr.zenczykowski@ifj.edu.pl}\\
{\em Division of Theoretical Physics}\\
{\em The Henryk Niewodnicza\'nski Institute of Nuclear Physics}\\
{\em Polish Academy of Sciences}\\
{\em Radzikowskiego 152,
31-342 Krak\'ow, Poland}\\
}
\maketitle
\begin{abstract}
We describe a MOND-related approach to natural scales of distance and mass, viewing it as a logical step
following Planck's modification of the Stoney system of units.
The MOND-induced scales are not based on the strength of any physical interaction (electromagnetic, gravitational, or  otherwise). Instead, they are specified by three physical constants
of a general nature that define the scales of action, speed, and acceleration, ie.
$h$ -- the Planck constant, $c$ -- the speed of light, and $a_M$ -- the MOND acceleration
constant. 
When the gravitational constant $G$ is added, two further distance
scales (apart from the size of the Universe) appear: the Planck scale and a nanometer scale that fits the typical borderline between the classical and the quantum descriptions.

\end{abstract}

\vfill
\end{titlepage}

\section{Introduction}
%\label{sec1}
In order to provide numerical description of a physical phenomenon 
one must choose units in which physical quantities are to be measured.
These units are often anthropocentric in spirit.
A question then emerges of how to introduce  units 
that --- instead of being 
specific to us --- would be related to some natural scales of the world we live in. 
Various systems of such units and related scales have been proposed. Probably the system most widely known is that of Planck, which constitutes a modification of an earlier proposal by Stoney.

\section{From Stoney to Planck}
The concept of natural units emerged around 1880 when G. J. Stoney \cite{Stoney,Tomilin} put forward his proposal on how to introduce non-anthropocentric units of mass ($m$), length ($l$), and time ($t$). According  to his idea, such units should be based on natural scales defined by some appropriately chosen universal constants of Nature. As the relevant constants he took $c$ -- the speed of light, $G$ -- the Newtonian constant of gravitation, and $e$ -- the electron charge. This choice is a testimony to Stoney's great physical intuition: at the time of his proposal it was not known that $c$ is the same for all inertial systems (and that it is the maximal speed 
with which physical bodies may move),  while the concept of the electron
was still in its infancy.\footnote{Actually, Stoney was the first to introduce this concept and to calculate the expected electron charge. \cite{Barrow}} Expressed in terms of these constants, Stoney's units were:
$m_S=\sqrt{e^2/G}$, $l_S= \sqrt{Ge^2/c^4}$, and $t_S = l_S/c$.\\

\noindent
In 1899, following his discovery of the quantum of action, Max Planck proposed \cite{Tomilin,Barrow} a different system of natural units, with the basic role ascribed not to the electron charge but to the quantum of action $h$. The Planck system may be considered an improved version of the Stoney system, with Planck's units related  to those of Stoney by
the dimensionless factor of $\sqrt{e^2/(hc)}$.  
With $G = 6.67 \times 10^{-8}~cm^3/(g~s^2)$, $c = 3 \times 10^{10}~cm/s$, and 
$h = 6.62 \times 10^{-27}~g~cm^2/s$,
the respective Planck scales
of mass, length, and time are:
$m_P=\sqrt{{hc}/{G}}=5.46 \times 10^{-5}~g$, 
  $l_P = \sqrt{hc/G^3} = 4.05 \times 10^{-33}~cm$, and $t_P=l_P/c = 1.35 \times 10^{-43}~s$.
The systems of Stoney and of Planck are
``more natural'' than various other systems, for they are not based 
on the arbitrary choice of an object (particle) to set the mass scale.
Planck did not offer any explanation as to the meaning of the miniscule size of his distance and time scales, considering  the presence of $G$ in his formulas
quite mysterious. 
It was only through Einstein's creation of general relativity (GR) that gravity was tied to the properties of space and time, thus making the appearance of $G$ in Planck's units  more
justifiable. Yet it took a long time before the current view started to dominate with its interpretation of $l_P$ and $t_P$ as the distance and time scales at which the quantum nature of space should become manifest. \\

\noindent
One may doubt whether (or in what sense) Planck's scales are ``natural'' for quantum gravity.
As clearly demonstrated by Meschini \cite{Meschini}, dimensional analysis is not an infallible tool that could provide us with absolutely sound information on the realm of the Unknown. In fact, it becomes reliable only when we know in advance which theory,  physical constants, and/or quantities should be chosen as relevant to a given problem.
\footnote{For example, in order to estimate the period of a pendulum at sea level one has to know first (from observations and theory) that the relevant quantities are the Earth gravity $g$ and the length of the pendulum, but not its mass. The connection with $G$ is then made through Newton's theory of gravitation, and requires the knowledge of the Earth mass and radius.} Yet we have no direct experimental handle on what happens at Planck's length (and time) scales.
Thus, it may well be that the distance scale at which the quantum nature of space becomes manifest is dynamical and much larger than $l_P$ \cite{Bojowald} (and the related mass scale --- much smaller than $m_P$). 
With space regarded as an attribute (or a derivative) of matter, it may even be claimed that it is physics at hadronic (or, more generally speaking, elementary particle) scales that should direct our ideas on the quantization of space \cite{Zen2018}.
In spite of such caveats, dimensional analysis certainly provides us with important hints on the Unknown. \\

\noindent
In this note we will use dimensional analysis and elaborate briefly on the observation that the systems of Stoney and of Planck are based on two types of universal constants:
those that place certain conditions or limits on the size of general physical quantities
(whose values were originally deemed arbitrary), and those that set the strength of specific interactions. 
Stoney's system uses one constant of the first type ($c$, the speed of light, which defines the upper limit on the velocity attainable by moving bodies, with the limit $1/c \to 0$ specifying the transition from the relativistic to the nonrelativistic regime), and two constants describing the strength of interactions considered to be all-important ($e$ and $G$). On the other hand,
Planck's system is based on two constants of the first type ($c$ and $h$, the latter defining the scale at which the quantum aspects of Nature become manifest, with the limit $h \to 0$ corresponding to the transition from the quantum to the classical regime),
and one constant of the second type ($G$).
Today we know that not all particles experience electromagnetic forces, but
all matter is subject to gravitational interactions.
Consequently, with quantum aspects being ubiquitous in Nature, Planck's system seems conceptually more fundamental than that of Stoney. \\

\section{Milgrom}
Now, following the discovery of the accelerated expansion of the Universe \cite{Riess} it appears that
the two constants relevant for classical gravity ($G$ and $c$) should be supplemented with a third (cosmological) constant of positive value: $\Lambda =1.11 \times 10^{-56}~cm^{-2}$. It seems therefore that the proper quantum theory of gravity (if gravity is to be quantized) should depend on all four dimensional constants $c$, $h$, $G$, and $\Lambda$ \cite{Smolin}. Planck's scales are then obtained if $\Lambda$ is not used.
If $\Lambda$ is taken into account but one of the remaining three constants is not,  additional mass scales appear.
The first is Wesson's mass $m_W =(h/c)\sqrt{\Lambda/3} = 1.34 \times 10^{-65}~g$, which is independent of $G$ and may be interpreted as the quantum of mass \cite{Wesson}. The second, defined
by the omission of $h$, is $m_U= (c^2/G)\sqrt{3/\Lambda}= 2.22 \times 10^{56}~g$, and is
of the order of the mass of the observable Universe. The third, independent of $c$ and given by the expression \footnote{This formula constitutes a variant of the Eddington-Weinberg relation \cite{EW,Zeldovich}.}
$m_N =\left((h^2/G)\sqrt{\Lambda/3}\right)^{1/3} = 0.34 \times 10^{-24}~g$, is of the order of nucleon
or pion mass~\cite{Zen2018}. 
The above observation that there is a set of four basic mass scales
($m_P$, $m_W$, $m_U$, and $m_N$) that can be obtained through dimensional analysis
 by omitting one of the four fundamental constants was made earlier in \cite{Burikham}. 
The interpretation of the meaning of these mass scales may depend on the associated (or underlying) physics. In his original paper introducing mass, distance and time  units $m_P$, $l_P$, and $t_P$, Planck did not suggest any  such interpretation  (eg. a  possible relevance of these scales for the idea of quantum gravity).   
In fact, for the consideration of the quantum nature of matter and space, 
with $m_N$ dependent on both $G$ and $\Lambda$ as well as on $h$, it may
 be claimed that 
$m_N$ is more appropriate than $m_P$ \cite{Zen2018}.
Connections of $m_N$ with other physical ideas can also be made.
For a cosmological interpretation see \cite{Capozziello}.
A yet different interpretation  is made in \cite{Burikham}, where $m_N$
is interpreted as the temperature ($k T_N \approx m_Nc^2$) of a black hole 
of mass $m_{BH} = m_U m_W/m_N$,
when the hole stops radiating
\footnote{It might be interesting to note that this temperature is of the order of Hagedorn
temperature in hadronic physics.}.  
\\

\noindent
The appearance of several gravity-related scales raises the question of ``the most natural'' ones. One expects that such scales should be defined with the help of $c$ and $h$, but the use of $G$ and/or $\Lambda$ (pertaining to the strength of the relevant interactions) may be questioned just as the use of $e$ was.
 Indeed, $G$ and $\Lambda$ do not seem to define different regimes of gravitational interactions in a way that would resemble the limiting roles played by $c$ and $h$ \cite{Milgrom2015}. 
Yet there are experimental 
hints that suggest the existence of another constant of a similarly limiting type. Such a constant ($a_M$), with the dimension of acceleration, was introduced by Milgrom \cite{Milgrom1983}
as a part of MOND, an unorthodox view on gravitational dynamics. 
\footnote{With different interpretations of mass scales apparently possible
(as discussed above for the case of $m_N$), below we  restrict our discussion to considerations that follow from the dimensional analysis only and, therefore, are not sensitive to the detailed physics possibly associated with $a_M$ 
(should such layer of physics be sought at all). In other words, 
we want to seek the general conclusions that follow from the replacement
of $G$ and $\Lambda$ by $a_M$, the latter being akin to 
 $h$ and $c$ (which two constants are not generally considered to result from some underlying physics). In our opinion, at this stage a particular ``microscopic'' explanation of $a_M$ (like that of \cite{Panpanich}) is of secondary interest only, as -- in analogy to $m_N$ -- various such connections may be expected.}\\

\noindent
The goal of MOND was 
to account for the deviation from the Newtonian expectations of the observed rotational velocities of stars and gas in the outer regions of galaxies \cite{TullyFisher}. While such discrepancies could be explained
through the introduction of  non-luminous
dark matter (as yet unobserved),
the novel idea of \cite{Milgrom1983}
was that Nature departs from Newtonian dynamics (and GR) for very low accelerations $a$, with $a_M$ 
marking the boundary between the 
Newtonian regime ($a > a_M$) and the MOND regime ($a < a_M$).
The difference between the two regimes consists in the adopted form of gravity-induced
acceleration. In the Newtonian domain this form is standard ($a = a_N \equiv GM/R^2$), 
while in the MOND domain it is taken to be $a = \sqrt{a_N~a_M}$. 
It is remarkable how elegantly many astrophysical problems can be solved with such a simple
modification of Newtonian dynamics. Indeed,
one finds that astrophysical data can be fitted with a single
universal value $a_M = (1.2 \pm 0.2)\times 10^{-8}~cm/s^2$. \\

\noindent
For the many successes of MOND and its possible superiority over the dark matter paradigm, see the relevant reviews, eg. \cite{McGaugh}.
These successes suggest to many physicists that 
MOND should not be viewed as an economical description of the effects of dark matter, but
that it touches on some --- still unknown but truly fundamental --- physics \cite{Smolin,Milgrom2015,Bernal}.
Accepting such views, $a_M$ should be regarded as a fundamental constant delineating different physical domains of universal gravity-induced forces in a way somewhat akin to that defined by $c$ and $h$ in other contexts. 
Thus, we argue here that the system of scales based on $c$, $h$, and $a_M$ should be viewed as a logical follow-up of  Planck's modification of the system of Stoney.\\

\noindent
Using $c$, $h$, and $a_M$, one defines MOND-related ``natural'' scales of length: 
\begin{equation}
\label{lM}
l_M = c^2/a_M \approx 7.5 \times 10^{28}~cm
\end{equation}
(which is of the order of the size of the observable Universe), time:
\begin{equation}
\label{tM}
t_M = c/a_M \approx 2.5 \times 10^{18}~s
\end{equation} 
(of the order of the Universe age),
and mass:
\begin{equation}
\label{mM}
m_M = h a_M/c^3 \approx 0.29 \times 10^{-65}~g.
\end{equation}
Proximity of $m_M$ and $m_W$ results from the  coincidence (as yet not understood)
between
the acceleration of the Universe expansion $a_{\Lambda} = c^2 \sqrt{\Lambda}$, and $a_M$:
\begin{equation}
a_{\Lambda} \approx 7.9 ~a_M.
\end{equation}
With $m_M$ incredibly small, the scales $l_M$, $t_M$, and $m_M$ seem to be removed from the anthropocentric scales even further than Planck's scales (though, for $l_M$ and $t_M$, in the opposite directions).
While the three scales of Eqs (\ref{lM},\ref{tM},\ref{mM}) are of course well known, we stress here their role as the final stage in a chain of
departures from the use of interaction-strength-defining constants. \\

\section{Milgrom and Planck}
In order to discuss some other distance and mass scales, relevant in the Newtonian/GR domain, 
we now reintroduce $\Lambda$ and $G$, and 
 consider products of $c$, $h$, $a_M$, $G$, and $\Lambda$,
raised to powers appropriate for the physical quantity in question, ie. 
\begin{equation}
c^{\alpha}h^{\beta}a_M^{\gamma}G^{\delta}{\Lambda}^{\epsilon}.
\end{equation}
With $\Lambda$ being of the order of
$a_M^2/c^4$, we accept that $\Lambda$ and $a_M$ are related, and
restrict our considerations to products involving
$c^{\alpha}h^{\beta}a_M^{\gamma}G^{\delta}$ only.
The appearance of $G$ permits the introduction of a dimensionless
number
\begin{equation}
\frac{G~h~a_M^2}{c^7} = 0.29 \times 10^{-122}.
\end{equation}
The new scales may then be obtained via the multiplication of $l_M$ and $m_M$ 
by arbitrary powers of the above number, ie. one may define
the length scales
\begin{equation}
l_{\delta} = \frac{c^2}{a_M} \left[\frac{G~h~a_M^2}{c^7}\right]^{\delta},
\end{equation}
and the mass scales:
\begin{equation}
m_{\delta} = \frac{h~a_M}{c^3}\left[\frac{G~h~a_M^2}{c^7}\right]^{-\delta}
\end{equation}
with $m_{\delta}l_{\delta} = h/c$.\\

\noindent
For the length scales, three particular values of $\delta$ 
may be singled out as special: when
$l_{\delta}$ 
is independent of $h$ (and of the product $Gh$) which occurs for $\delta=0$, when it is independent of $c$ 
(for $\delta = 2/7$), and when it is independent of $a_M$ ($\delta = 1/2$). 
These three cases are somewhat similar to one another with one of the fundamental limiting constants ($h$, $c$, $a_M$) not used. 
Then, $l_0$ is equal to $l_M=c^2/a_M$, while $l_{1/2}$ to $l_P=\sqrt{(Gh)/c^3}$, 
with both of these scales regarded as scales of important transitions.
For $l_{2/7}$
we have
\begin{equation}
\label{l27}
l_{2/7} = \left[\frac{G^2h^2}{a_M^3} \right]^{1/7} = (7.32 \pm 0.5)\times 10^{-7} cm.
\end{equation}
The latter value is of the order of the width of a haemoglobin molecule or the size of the gate length of a 16 nm processor \cite{length}, and is mere two orders of magnitude above Bohr radius $r_B = \hbar^2/(m_ee^2) = 5.29 \times 10^{-9}~cm$. The atomic sizes are in the range of $(0.1-0.5) \times 10^{-7}~cm$, an order of magnitude below (\ref{l27}). Thus, our new formula (\ref{l27}) fits another important scale --- the distance scale corresponding to the transition from the classical to the quantum description of ordinary matter. This does not seem to be just a coincidence: from the whole (and very wide) range of possible distance scales in between the Planck and the Universe scales,
% $(l_{1/2},l_0) \approx (4\times 10^{-33}, 7 \times 10^{28})~cm$, 
prescription (\ref{l27})
%$l_{2/7}$ 
selects the value of a typical (nanometer) classical-to-quantum transition scale. With atomic sizes being of electromagnetic origin, the above coincidence suggests an unknown link between electromagnetism and gravity.   
\\

\noindent
Similarly, for the mass scales there are four particular values of $\delta$ that may be singled out. 
These are the cases when $m_{\delta}$ is independent of $G$ (i.e. $\delta = 0$), of $c$ 
($\delta = 3/7$), of $a_M$ ($\delta =1/2$), and of $h$ ($\delta =1$).   With $m_{1/2}=\sqrt{hc/G}$ and $m_1=c^4/(Ga_M)$ being, respectively, the Planck and the Universe masses, one then expects $m_{3/7}$ to be also of some significance.
One finds
\begin{equation}
\label{m37}
m_{3/7} = \left[\frac{h^4~a_M}{G^3}\right]^{1/7} = 0.96 \times 10^{-13}~g,
\end{equation}
which is several orders of magnitude both above the atomic mass ($10^{-22}~g$) and below Planck's mass ($5\times10^{-5}~g$).
For comparison, a typical bacterium has a mass of $10^{-12}~g$, while viruses have masses of the order of $10^{-17} - 10^{-14}~g$ \cite{bacteriumvirus}. As masses correspond to objects living in 3D space, comparison with the case of lengths should be made only after taking cubic roots of the relevant quantities.
\\

\noindent
In the derivation of the length and mass scales of Eqs (\ref{l27},\ref{m37}), 
 we used $G$, $a_M$, and $h$ while discarding $c$.  Had we replaced $a_M$ with 
$c^2\sqrt{\Lambda}$, and
used $G$, $\Lambda$, and $h$ (i.e. omitting $c$), we would have obtained the nucleon mass scale  $m_N =\left((h^2/G)\sqrt{\Lambda/3}\right)^{1/3}$ and the Universe scale $l_U = 1/\sqrt{\Lambda}$ only. Thus, the use of a limiting constant $a_M$ in place of the cosmological constant $\Lambda$ has been essential in our derivation of the nanometer 
transition scale.\\

\noindent
Given the above comparisons it is tempting to suspect that a typical transition between the classical regime of the large and the quantum regime of the small occurs for length scales which may be defined with the help of $h$, $G$, and $a_M$ (ie. without $c$). We refrain from speculations as to the possible deeper  meaning of this observation. 
Yet it looks as if various macro- and microscopic transition scales
 were more closely related than usually believed.

\vfill

\vfill

\end{document}